\documentclass[a4,paper,11pt]{article}
\usepackage{booktabs}
\usepackage{epsfig}
\usepackage{graphicx}
\usepackage{rotating}
\usepackage{float}
\usepackage{subfigure}
\usepackage{amsmath,amsmath,mathrsfs}
\allowdisplaybreaks
\usepackage{cite}
\usepackage{mathrsfs}
\usepackage{multirow}
\headheight\baselineskip\headsep2\baselineskip\textwidth200mm
\advance\textwidth -2in\textheight280mm\advance\textheight-2in
\usepackage{amsfonts}
\usepackage{graphicx}
\usepackage{amssymb}
\usepackage{graphicx}
\usepackage{amsmath}
\usepackage{lscape}
\usepackage{epsfig}
\usepackage{amsfonts,amssymb,amsmath,lscape}
\usepackage{colortbl}
\usepackage{epsfig}
\usepackage{graphicx}
\usepackage{epstopdf}
\usepackage[]{algorithm2e}
\usepackage{algpseudocode}
\usepackage{amsthm}
\usepackage{amsfonts}
\usepackage{amssymb}
\usepackage{graphicx}
\usepackage{amsmath}
\usepackage{lscape,color}

\definecolor{c30}{rgb}{0,0,1}

\newtheorem{Example}{Example}
\newtheorem{Remark}{Remark}

\usepackage{tikz}

\begin{document}

\title{\bf \LARGE An Algorithm for computing the t-signature of two-state networks}
\author{
\textbf{M. Siavashi\footnote{Department of Computer Science, Engineering, and IT, School of Electrical and Computer Engineering, Shiraz University,  Shiraz 71454, Iran (E-mail: m.siavashi@cse.shirazu.ac.ir)}, \ \ and \ \ S. Zarezadeh\footnote{Department of Statistics, Shiraz University, Shiraz 71454, Iran  (E-mail: s.zarezadeh@shirazu.ac.ir)}}\\
}
\date{}
\maketitle
\begin{abstract}

\begin{small}
Due to the importance of {\it signature} vector in studying the reliability of networks, some methods have been proposed by researchers to obtain the signature.  
The notion of signature is used when at most one link may fail at each time instant.
It is more realistic to consider the case where non of the components, one component or more than one component of the network may be destroyed at each time. Motivated by this, the concept of t-signature has been recently defined to get the
reliability of such a network.  The t-signature  is a probability vector and depends only on the network structure.  In this paper, we propose an algorithm to  compute  the  t-signature. The performance of the proposed algorithm is evaluated for some networks.\\

{\bf Keywords:} Network reliability, BFS algorithm, signature, binomial distribution.
\end{small}
\end{abstract}
\section{Introduction}
Network reliability modeling has been widely investigated in the literature. A network is defined as a collection of nodes, and links in which some particular nodes are considered as {\it terminals} set. Rail stations, telecommunication centers, and computers are examples of nodes, and rail ways,  communication channels, and the cables between the computers are examples of links. The states of network are usually defined  based on the connection between the terminals. In this paper, we assume that the nodes are absolutely reliable and the links are subject to failure.  The network has two states: {\it up}, and {\it down} and the network state may change in the process of links failure. It should be mentioned that a coherent system can be considered as a two-state network with two terminals.

There are different models to get the reliability of networks.
One of these models, which has been considerably explored, represents the reliability of network as a mixture of the reliability of ordered lifetimes of links; see \cite{saman} and \cite{Gert-2011}.  Let $T$ be the lifetime of a network having $n$ links with independent and identically distributed (i.i.d.) lifetimes $T_1,\dots,T_n$. Under the assumption that there are not ties between the occurrence times of links failure, $P(T_i=T_j)=0$, $i\neq j$, the network reliability is written as
\begin{align}\label{model-saman}
P(T>t)=\sum_{i=1}^n s_i P(T_{i:n}>t), \quad t>0,
\end{align}
where $T_{i:n}$ is the $i$th ordered lifetime of links and $s_i=P(T=T_{i:n})$ , $i=1,\dots,n$. The probability vector ${\bf s}=(s_1,\dots,s_n)$ is  called {\it signature} which does not depend on the random mechanism of links failure and  is only determined based on the network structure.
Similar mixture representation is hold for the reliability of network when the links of network fail based on a counting process; see, \cite{Gert-2011} and \cite{za-as-ieee}.
In recent years, many researchers have been investigated the modeling of network reliability based on signature and its variants  under various schemes and applications.  See, among others,  \cite{Eryilmaz-2018}, \cite{saman-2007}, \cite{Coolen and Coolen-Maturi (2015)}, \cite{Zarezadeh et al. (2018)}.

The concept of signature has also combinatorially definition given in \cite{Gert-2009} as follows:
Consider a network with $n$ links and let $\pi=(e_{i_1},e_{i_2},\ldots,e_{i_n})$ denote a permutation of the ordinal number of network links. Assume that all links are in up state and by moving from left to right of permutation, turn the state of each link from up  to down.   By the assumption that all permutations are equally probable,  the signature is defined as ${\bf s}=(s_1,\dots,s_n)$ where
\begin{align}
 s_i=\frac{n_i}{n!},\qquad i=1,\dots, n,\label{signa}
\end{align}
where $n_i$ is the number of permutations in which the failure of  the $i$th link causes the change of network state from up to  down.
It is difficult to compute the signature of a network with a large number of
links. Two basic formulas were given in \cite{Da et al. (2012)} to compute the signature of a system which can be decomposed into two subsystems.
Da et al. \cite{Da et al.(2014)} derived some formulas for computing the signature of a $k$-out-of-$n$ system consisting of $n$ modules.  Marichal and Mathonet \cite{Marichal and Mathonet (2013)} proposed a method to get the signature of system using the diagonal section of the reliability function via derivatives.  An algorithm was suggested in \cite{Gert-2009} to calculate the signature of two-state networks. Da et al. \cite{Da2018} gave an efficient algorithm for computing the signatures of systems with exchangeable components.

As mentioned before, the notion of signature is applicable when it is not possible to fail more than one link at each time instant; see, e.g.,  \cite{Gert-2009}, \cite{saman-2007}. It is more realistic to consider the case where more than one link may fail at each time. Motivated by this, Zarezadeh et al. \cite{zaa} studied the reliability of two-state networks under the aforementioned assumption. 
They considered a network which is subject to shocks and each shock may lead to links failure.  The network finally fails  by one of these  shocks. Further, it was assumed that $N(t)$ denotes the number of links  that fail up to time $t$, and $T$ is the network lifetime.  Under the  assumption that the process of occurrence of the shocks is independent of the number of failed links, it was shown that
\begin{align}\label{M}
P(T>t)=&\sum_{i=1}^n s_i^{\tau} P(N(t)\leq i-1).
\end{align}
where ${\bf s}^{\tau}=(s_1^{\tau},s_2^{\tau},\dots,s_n^{\tau})$ is called {\it t-signature} vector.  The t-signature has the following combinatorial definition:\\ If there is possibility of failure of more than one link at each time instant, then the way of the order of links failure is different from ordinal permutation applied in the definition of signature. All ways of links failure are obtained in two stages: first we obtain all partitions of $\{1,\dots,n\}$ and then all permutations of the elements of each partition are considered. Therefore, the number of ways of order of links failure, denoted by $n^*$, has been obtained as
\begin{align}\label{n^*}
n^*=\sum_{j=1}^{n}\sum_{k=0}^{j}{j \choose k} (-1)^k (j-k)^n;
\end{align}
see Lemma 1 of \cite{zaa}.
Let the discrete  random variable $M$ denote the minimum number of links whose failures cause to fail the network in a way of links failure order. Clearly, $M$ takes values on $\{1,2,\dots,n\}$. Suppose $n_i$ is the number of ways of the order of links failure in  which $M=i$. Assuming that all  ways of  order of links failure are equally likely, the  t-signature  vector associated to the network is defined as ${\bf s}^{\tau}=(s_1^{\tau},\dots, s_n^{\tau})$  where
\begin{align*}
s^{\tau}_i=\frac{n_i }{n^*}, \qquad i=1,\dots,n.
\end{align*}
It is clear that the t-signature depends only on the structure of the network and is free of the random mechanism of links failure. It should be mentioned that the t-signature converts to signature, defined in \eqref{signa}, when only usual permutations of links number are considered as all ways of order of links failure.  An extension of t-signature to networks with three states is given in \cite{AZ18}.
Due to the importance of t-signature in exploring the reliability of two-state networks,  the aim of this paper is to propose an algorithm for computing  the t-signature vector.   For some networks, the performance of the algorithm is examined.

\section{The proposed algorithm}
Because of the importance of t-signature in analyzing the reliability and the failure time of networks, we give an algorithm for getting the t-signature of networks. To this, we first give the following remark which has a key role in the proposed  algorithm for computing the minimum number of links whose failure causes  the network failure in each  way of links failure order, $M$.
\begin{Remark}\label{remark}
A minimal path set is a minimal set of links whose working ensures the function of network. {To fail the network, it is necessary to disconnect all corresponding minimal path sets.} 
\end{Remark}

Let us first  introduce the following notations.
\subsubsection*{Notations:}
\begin{tabular}{ll}
 $n$ & The number of links of the network.\\
 $n^*$ & The number of all ways of order of links failure\\
 {partitions}: & All partitions of the given set \\
 {part}: & An element of {\it partitions} \\
 results: & An array to save the result of algorithm \\
 all\_orders: & An array with $n^*$ elements which denotes all ways of order of links failure\\
 order & An element of {\it all\_orders}. Note that {\it order} is a vector whose elements \\
  & are subsets of $\{1,\dots,n\}$. \\
 source\_node & The source node (terminal)\\
 destination\_node & The destination node (terminal)\\
 removed\_links: & An array to save the links which have been removed\\
 failed\_set: & The element of {\it order} which fails. \\
 $M$ & The minimum number of links whose failure causes to fail\\
 & the network in each order.\\
\end{tabular}
\\
\\

From the definition of t-signature, for computing the t-signature, we need to find all ways of order of links failure. Then we first give an algorithm for this purpose.

\begin{algorithm}[H]
\DontPrintSemicolon 
\KwIn{The set of network links}
\KwOut{All ways of order of links failure}
initialization\;
all\_orders $\gets$ [ ]\;
    \For{part in partitions(links):}{ 
        all\_orders.push(permutations(part))} 
    \Return{all\_orders}\;
\caption{An algorithm to derive all\_orders}
\end{algorithm}

Consider a network represented by a graph ${\mathcal G}$ in which the nodes are absolutely reliable and the links are subject to failure.  Two nodes of  network are selected as the source and the destination nodes (terminals set) of the network.
\\
The following algorithm gives the t-signature corresponding to the network.\\
\begin{algorithm}[H]
\DontPrintSemicolon 
\KwIn{The graph of network and its terminals set.}
\KwOut{The t-signature of network, ${\bf s}^\tau=(s_1^\tau,\dots,s_n^\tau)$.}
initialization\;
results $\gets$ [ ]\;
    \For{order in all\_orders:}{ 
        results.push(calculate\_m(order))} 
    \For{$i$ in range(1,$n$):}{ 
        $n_i$ $\gets$ count(results, $i$)\;
        $s_i^\tau \gets \dfrac{n_i}{n^*}$\;
        {{\bf Print} {$s_i^\tau$}}}\;
\caption{The algorithm for computing the exact value of t-signature}\label{algor-main}
\end{algorithm}

In Algorithm \ref{algor-main}, ``{\it calculate\_m}'' calculates the amount of $M$ for each order of links failure. In fact, computation of $M$ is the main part of our algorithm which we give the following algorithm for its computation.\\
\begin{algorithm}[H]
\DontPrintSemicolon 
\KwIn{A way of order of links failure ({\it order})}
\KwOut{The amount of $M$ for {\it order}, {\it calculate\_m}({\it order})}
initialization\;
    M $\gets$ 0\;
    removed\_links $\gets$ [ ]\;
    \For{failed\_set in order:}{
        \eIf{graph.has\_route(source\_node, destination\_node, {append(removed\_links, failed\_set)}):}{
          $M \gets M +$length(failed\_set)\;
          removed\_links $\gets$ {removed\_links+failed\_set}}{
          $M \gets M + $find\_min\_m(source\_node, destination\_node, removed\_links, failed\_set)\;
          break
}}
    \Return{M}\;
\end{algorithm}

\begin{algorithm}[H]
\DontPrintSemicolon 
\KwIn{source\_node, destination\_node, removed\_links, failed\_set}
\KwOut{The amount of $m$}
initialization\;
     m=0\;
    \While{graph.has\_route(source\_node, destination\_node, removed\_links)}
    {path=graph.find\_path(source\_node, destination\_node, removed\_links)\;
    \For{link in path:}{
    \If{link in failed\_set}
    {append(removed\_links, link)}
    }
    $m\gets m+1$\;}
    \Return{m}\;
\end{algorithm}
Function ``{\it graph.hase\_route}'' explores the existence of path between {\it source\_node} and {\it destination\_node} using DFS or BFS algorithm when all links in {\it removed\_links} and {\it failed\_set} are removed. 
For the network after removing all links in {\it removed\_links}, function {\it find\_min\_m}, based on Remark \ref{remark}, explores the minimum number of links of {\it failed\_set} whose failure cause the  failure of network. 
The function ``graph.find\_path'' finds a path between the source and destination nodes with respect to removed\_links.

It is remarkable that the proposed algorithm can be also applied for networks with more than two terminals. If the network is defined to be in up state if and only if all terminals are connected, the function ''$graph.has\_route$'' should explore the {existence of a path from a terminal to all other terminals which finally connects all terminals} and the function {\it find\_min\_m} explores the minimum number of links of {\it failed\_set} whose failure cause to disconnect the terminals of network.

As seen in Table \ref{tab-n*}, the number of $n^*$ is very large even for $n=9$. Then we need to get the approximate t-signature for large value of $n^*$. To this, we select the samples using the method of probability proportional to size (PPS) sampling 
{in which the probability of selecting an order from partitions with $k$ parts is equal to ${m_k}/{n^*}$ where $m_k$ is the size of total number of ordered partitions with $k$ parts.}
\begin{table}[ht]
\caption{The amount of $n^*$ for different values of $n$.}\label{tab-n*}
\begin{center}
\small
\begin{tabular}{cll}
  \hline
  $n$ & $n!$ & $n^*$ \\
  \hline
  2 & 2 & 3 \\
  3 & 6 & 13 \\
  4 & 24 & 75 \\
  5 & 120 & 541 \\
  6 & 720 & 4,683 \\
  7 & 5,040 & 47,293 \\
  8 &  40,320 & 545,835\\
  9 &  362,880 & 7,087,261\\
  10 & 3,628,800 &  102,247,563\\
  11 &  39,916,800 & 1,622,632,573\\
  12 & 479,001,600 & 28,091,567,595\\
  \hline
  \end{tabular}
\end{center}
\end{table}

\section{Examples}
In this section, we examine the algorithm for some networks.
The computer program is developed in Python v2.7.10. To run this
program we use an intel core i5-4200U processor 1.6 GHz and 8 GB RAM under Windows 8 64-bit.

\begin{Example} \em
Consider the network with graph as depicted in Figure \ref{fig-8links} with terminals set $\mathscr{T}=\{b,c,d\}$. Let the links be subject to failure and the network be defined to be in up state if and only if all terminals are connected. Using the proposed algorithm, the t-signature vector is exactly obtained, by exploring all $n^*=7087261$ possible permutations, as follows
\[{\bf s}^\tau=(0,0.1030962,0.2788933,0.4374931,0.1512359,0.0292814,0,0,0).\]

\tikzstyle{block} = [shape=circle,draw,fill=blue!20,minimum size=1.5em]
\begin{figure}[H]
\begin{minipage}{.5\linewidth}
\begin{center}
\begin{tikzpicture}[node distance = 5 cm]
  \tikzset{LabelStyle/.style =   {scale=1, fill= white, text=black}}
  \node[block,scale=0.95] at (-1.7,0) (a) {$a$};
  \node[block,scale=0.9] at (-1,-1.5) (b) {$b$};
  \node[scale=1,block] at (-0.8,1.2) (c) {$c$};
  \node[block] at (2,-1.5) (e) {$e$};
  \node[block,scale=0.9] at (0.2,-0.5) (d) {$d$};
  \node[block,scale=0.85] at (1.5,1.1) (f) {$f$};
\node[block,scale=0.85] at (2.2,0.1) (g) {$g$};
  \draw[semithick](a) to  (b);\draw[semithick](c) to  (d);\draw[semithick](a) to  (d);\draw[semithick](a) to  (c);\draw[semithick](b) to  (e);\draw[semithick](c) to  (f);\draw[semithick](g) to  (d);\draw[semithick](g) to  (f);\draw[semithick](e) to  (g);
  \end{tikzpicture}
\caption{\small The network with 9 links and 7 nodes.}
\label{fig-8links}
\end{center}
\end{minipage} \qquad
\begin{minipage}{.5\linewidth}
\begin{center}
\begin{tikzpicture}[node distance = 5 cm]
  \tikzset{LabelStyle/.style =   {scale=1, fill= white, text=black}}
  \node[block,scale=0.95] at (-2.5,1) (n1) {$a$};
  \node[block,scale=0.9] at (-2.5,-1) (n2) {$b$};
  \node[scale=1,block] at (-1,1.5) (n3) {$c$};
  \node[block] at (-1,-1.5) (n4) {$e$};
  \node[block,scale=0.9] at (0.5,0) (n5) {$d$};
  \node[block,scale=0.85] at (0.5,0.8) (n6) {$f$};
  \node[block,scale=0.95] at (0.5,-0.8) (n7) {$g$};
\draw[-](n1)--(n2); \draw[-](n1)--(n3); \draw[-](n1)--(n4); \draw[-](n2)--(n3);\draw[-](n2)--(n4);
\draw[-](n3)--(n5);\draw[-](n5)--(n4);\draw[-](n3)--(n6);\draw[-](n3)--(n7);\draw[-](n6)--(n4);\draw[-](n7)--(n4);
  \end{tikzpicture}
 \caption{\small The network with 11 links and 7 nodes.}
\label{fig-systemmm}
\end{center}
\end{minipage}
\end{figure}
\end{Example}

\begin{Example}\label{example-1}
\em Consider the network with the graph as depicted in Figure \ref{fig-systemmm}. This network has {\color{blue}$7$} nodes and {\color{blue}$11$} links in which the network is in up state if and only if there is a path between nodes $a$ and $d$.
As seen in Table \ref{tab-n*}, $n^*=1,622,632,573$. To explore the accuracy of approximation, we get both exact t-signature  and approximated signature. The result show that the accuracy of approximation is well even for $n=30000$.

The algorithm can be coded using parallelism methods.
Tables \ref{exp-g} and \ref{exp-g1} represent the exact value of t-signature and approximated t-signature with sample sizes $n=10^5,~10^6,~10^7,~10^8$, respectively, without parallelism method and using multi-threaded programming.
\begin{table}[H]
\caption{\small The exact and approximated t-signature (single thread) in Example \ref{example-1}}\label{exp-g}
\begin{center}
\small
\begin{tabular}{cccccc}
  \hline
$i$ & $10^5$	&	$10^6$	&	$10^7$	&	$10^8$ & Exact value\\
\hline
1  & 0.0      	&	0.0	    & 	0.0   	&	0.0     & 0.0 \\
2  & 0.02767	&	0.02775	&	0.02767	&	0.02765 & 0.02621 \\
3  & 0.05242	&	0.05292 &	0.05283 &	0.05277 & 0.05111 \\
4  & 0.08791	&	0.08852 &	0.08844 &	0.08861 & 0.08714 \\
5  & 0.15135	&	0.15222	&	0.15223	&	0.15229 & 0.15056 \\
6  & 0.23675	&	0.23564	&	0.23536	&	0.23548 & 0.23622 \\
7  & 0.21290	&	0.21272 &	0.21295	&	0.21284 & 0.21530 \\
8  & 0.13491	&	0.13509	&	0.13520	&	0.13520 & 0.13705 \\
9  & 0.06933	&	0.06920	&	0.06931	&	0.06912 & 0.07020 \\
10 & 0.02677    &	0.02593	&	0.02601	&	0.02602 & 0.02621 \\
11 & 0.0        &	0.0	    &	0.0   	&	0.0     & 0.0     \\
\hline
Time (seconds) & 7.60277	&	75.84628& 774.67163	& 7686.12945 & 43246.43322 \\
\hline
\end{tabular}
\end{center}
\end{table}

\begin{table}[H]
\caption{\small The exact and approximated t-signature (16 threads) in Example \ref{example-1}}\label{exp-g1}
\begin{center}
\small
\begin{tabular}{ccccc}
  \hline
$i$ & $10^5$	&	$10^6$	&	$10^7$	&	$10^8$	\\ \hline
1   & 0.0	        &	0.0	    &	0.0	    &	0.0	    \\
2   & 0.02763	&	0.02763	&	0.02765	&	0.02764	\\
3   & 0.05343	&	0.05264 &	0.05279	&	0.05275	\\
4   & 0.08688	&	0.08844 &	0.08866	&	0.08861	\\
5   & 0.15034	&	0.15243 &	0.15223 &	0.15233	\\
6   & 0.23716	&	0.23599 &	0.23559	&	0.23550	\\
7   & 0.21419	&	0.21330 &	0.21272 &	0.21291	\\
8   & 0.13469	&	0.13459 &	0.13530	&	0.13515 \\
9   & 0.06960	&	0.06901 &	0.06904	&	0.06908 \\
10  & 0.02609	&	0.02597 &	0.02603	&	0.02602	\\
11  & 0.0	    &	0.0	    &	0.0	    &	0.0	    \\
\hline
Time (seconds) & 3.87203	&  27.03228 & 230.24221	&	2234.29665\\
\hline
\end{tabular}
\end{center}
\end{table}

\end{Example}

\begin{Example}\label{cities}\em
Consider the network of European cities with graph depicted in Figure \ref{COST239 EON}. This network has $11$ nodes and $26$ links in which the network is in up state if and only if there is a path between nodes $\mathrm{PAR}$ and $\mathrm{COP}$.
Table \ref{cities-table} shows the approximated t-signature for given sample sizes. As shown in Table \ref{cities-table} the sampling method is accurate enough and more efficient to be used instead of running the algorithm for exact answer ($n^*=4,002,225,759,844,168,492,486,127,539,083$) for this network. The values in Table \ref{cities-table} are average of three runs.
\begin{figure}[h]
\small
\centering
\begin{tikzpicture} [thick, scale=1.2]
 \begin{scope}[auto,%
  every node/.style={scale=0.7,draw,font=\large, fill=gray!40,circle,minimum size=4em},node distance=2cm]
 \node[semithick,scale=0.9, shape = circle,draw, fill= white, text=black, inner sep =2pt, outer sep= 0pt, minimum size= 5 pt](8) at (0,0.1) {LUX};
  \node[semithick,scale=0.85, shape = circle,draw, fill= white, text=black, inner sep =2pt, outer sep= 0pt, minimum size= 5 pt](7) at (-0.5,1) {AMS};
   \node[semithick,scale=0.9, shape = circle,draw, fill= white, text=black, inner sep =2pt, outer sep= 0pt, minimum size= 5 pt](9) at (-1.2,0.3) {BRU};
  \node[semithick,scale=0.85, shape = circle,draw, fill= white, text=black, inner sep =2pt, outer sep= 0pt, minimum size= 5 pt](1) at (-1.5,-1) {PAR};
  \node[semithick,scale=0.75, shape = circle,draw, fill= white, text=black, inner sep =2pt, outer sep= 0pt, minimum size= 5 pt](10) at (-2,1.5) {LON};
  \node[semithick,scale=0.95, shape = circle,draw, fill= white, text=black, inner sep =2pt, outer sep= 0pt, minimum size= 5 pt](3) at (1.2,-1.2) {ZUR};
    \node[semithick,scale=0.9, shape = circle,draw, fill= white, text=black, inner sep =2pt, outer sep= 0pt, minimum size= 5 pt](2) at (1.5,-2) {MIL};
      \node[semithick,scale=0.9, shape = circle,draw, fill= white, text=black, inner sep =2pt, outer sep= 0pt, minimum size= 5 pt](4) at (2,-0.5) {PRA};
        \node[semithick,scale=0.9, shape = circle,draw, fill= white, text=black, inner sep =2pt, outer sep= 0pt, minimum size= 5 pt](6) at (3,1) {BER};
          \node[semithick,scale=0.9, shape = circle,draw, fill= white, text=black, inner sep =2pt, outer sep= 0pt, minimum size= 5 pt](5) at (3.5,-1.6) {VIE};
          \node[semithick,scale=0.75, shape = circle,draw, fill= white, text=black, inner sep =2pt, outer sep= 0pt, minimum size= 5 pt](11) at (1.5,3) {COP};
          \end{scope}
\draw[semithick](1) to  (8);
\draw[semithick](8) to  (7);
\draw[semithick](2) to  (1);
\draw[semithick](8) to  (9);
\draw[semithick](8) to (4);
\draw[semithick](11) to (10);
\draw[semithick](10) to (9);
\draw[semithick](10) to (1);
\draw[semithick](6) to (5);
\draw[semithick](7) to (9);
\draw[semithick](8) to (3);
\draw[semithick](11) to (7);
\draw[semithick](11) to (6);
\draw[semithick](11) to (4);
\draw[semithick](2) to (9);
\draw[semithick](7) to (10);
\draw[semithick](7) to (6);
\draw[semithick](1) to (9);
\draw[semithick](2) to (3);
\draw[semithick](2) to (5);
\draw[semithick](4) to (3);
\draw[semithick](4) to (5);
\draw[semithick](4) to (6);
\draw[semithick](1) to (6);
\draw[semithick](5) to (3);
\draw[semithick](1) to (3);
     \end{tikzpicture}
\caption{\small COST239 EON topology with 11 nodes and 26 links}
\label{COST239 EON}
\end{figure}
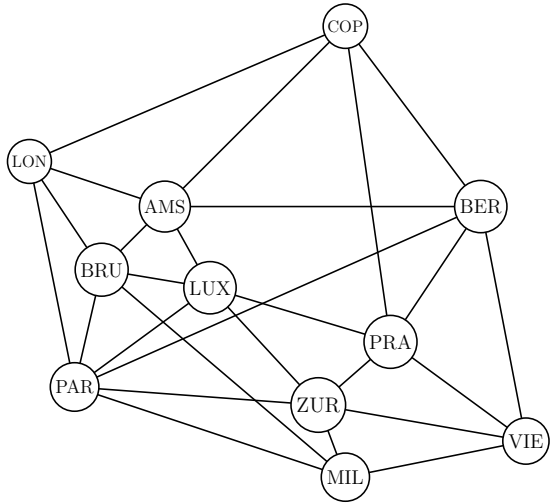

\begin{table}[H]
\caption{\small The approximated t-signature in Example \ref{cities}}\label{cities-table}
\begin{center}
\small
\begin{tabular}{ccccc}
  \hline
  $i$  & $N=10^{5}$ & $N=10^{6}$ & $N=10^{7}$ & $N=10^{8}$\\
  \hline
1	& 0.0	    & 	0.0	        &	0.0	        &	0.0\\
2	& 0.0	    &	0.0     	&	0.0	        &	0.0\\
3	& 0.0	    &   0.0 	    &	0.0	        &	0.0\\
4	& 0.000123	&	0.000096    &	0.000108 	&	0.000107\\
5	& 0.000403	&   0.000358	&   0.000366	&	0.000370\\
6	& 0.00077	&   0.00083	    &	0.000881	&	0.000868\\
7	& 0.001817	&	0.001743	&	0.001713	&	0.001716\\
8	& 0.003137	&	0.003054	&	0.003083	&	0.003077\\
9 	& 0.00521	&	0.00516	    &	0.005180	&	0.005183\\
10 	& 0.00853	&	0.008430	&	0.008462 	&	0.008481\\
11 	& 0.013697	&	0.013726	&	0.013630	&	0.013657\\
12 	& 0.022213	&	0.021826	&	0.021973	&	0.021973\\
13 	& 0.035277	&	0.035434	&	0.035208	&	0.035222\\
14 	& 0.055277	&	0.055687	&	0.055619 	&	0.055667\\
15 	& 0.083723	&	0.084352	&	0.084427 	&   0.084447\\
16 	& 0.116567	&	0.117488	&	0.117858	&	0.117803\\
17 	& 0.143217	&	0.143312	&	0.142924	&	0.142872\\
18 	& 0.143917	&	0.143148	&	0.143169	&	0.143175\\
19 	& 0.124207	&	0.122828	&	0.122804	&	0.122817\\
20 	& 0.094087	&	0.093965	&	0.093951	&	0.093849\\
21 	& 0.06492	&	0.065301    &	0.065438	&	0.065459\\
22	& 0.041783	&   0.041897	&	0.041940	&	0.041978\\
23	& 0.024777	&   0.024554 	&	0.024533	&	0.024496\\
24	& 0.012253	&   0.012403	&	0.012319	&	0.012354\\
25	& 0.004097	&   0.004406 	&	0.004414 	&	0.004428\\
26	& 0.0	    &   0.0	        &   0.0	        &	0.0\\
\hline
\end{tabular}
\end{center}
\end{table}

Also the t-signature of EON network with terminals set $\mathscr{T}=\{\mathrm{LON, BER, MIL}\}$ and $N=10^8$ samples is given in Table \ref{cities-taable}.
\begin{table}[H]
\caption{\small The approximated t-signature in Example \ref{cities}}\label{cities-taable}
\begin{center}
\small
\begin{tabular}{cccc}
\hline
$i$  & $s_i^\tau$ & $i$  & $s_i^\tau$\\
 \hline
1  &0.0        & 14 &0.0918775\\
2  &0.0        & 15 &0.1277313\\
3  &0.0        & 16 &0.1600846\\
4  &0.0002034  & 17 &0.1672263\\
5  &0.0007383  & 18 &0.1311279\\
6  &0.0017856  & 19 &0.0809734\\
7  &0.0035858  & 20 &0.0416957\\
8  &0.0064395  & 21 &0.0182939\\
9  &0.0107812  & 22 &0.0066694\\
10 &0.0173560  & 23 &0.0018301\\
11 &0.0270999  & 24 &0.0002743\\
12 &0.0415685  & 25 &0.0\\
13 &0.0626574  & 26 &0.0\\
\hline
\end{tabular}
\end{center}
\end{table}

\end{Example}

\bibliographystyle{amsplain}

\end{document}